\documentclass{IEEEtran}

\usepackage{graphics,graphicx} 
\usepackage{epsfig} 

\usepackage{amsmath}
\usepackage{amssymb}
\usepackage{tikz}
\usepackage{amsthm}
\usepackage{comment}
\usepackage[export]{adjustbox}
\usepackage{mathcomSTEP}
\usepackage{subcaption,setspace}
\usepackage[utf8]{inputenc}

\usepackage{tikz}
\usetikzlibrary{shapes,arrows,backgrounds,positioning}

\allowdisplaybreaks

\theoremstyle{remark}

\newtheorem{definition}{Definition}

\usepackage{tikz}
\usetikzlibrary{positioning}
\usetikzlibrary{arrows,fit}
\usetikzlibrary{decorations.pathreplacing}

\usepackage{epstopdf}
\usepackage{pgfplots,setspace}

\tikzstyle{int}=[draw, fill=blue!10, minimum height = 1cm, minimum width=1.5cm,thick ]
\tikzstyle{sum}=[circle, fill=blue!10, draw=black,line width=.5 pt,minimum size = 0.05 cm, thin ]
\tikzstyle{joint} = [draw, circle, minimum size=1em]

\begin{document}

\title{
On the Capacity of the Dirty Paper Channel with Fast Fading and Discrete Channel States
}

\author{%
\IEEEauthorblockN{%
Stefano Rini \IEEEauthorrefmark{1} and Shlomo  Shamai (Shitz) \IEEEauthorrefmark{2} \\}

\IEEEauthorblockA{%
\IEEEauthorrefmark{1}
National Chiao-Tung University, Hsinchu, Taiwan\\
E-mail: \texttt{stefano@nctu.edu.tw} }

\IEEEauthorblockA{%
\IEEEauthorrefmark{2}
Technion-Israel Institute of Technology,  Haifa, Israel \\
E-mail: \texttt{sshlomo@ee.technion.ac.il}
}

\thanks{
The work of S. Rini was funded by the  Ministry Of Science and Technology (MOST) under the grant 103-2218-E-009-014-MY2.
The work of S. Shamai was supported by the Israel Science Foundation (ISF).
}
}

\maketitle

\author{%
\IEEEauthorblockN{%
Stefano Rini \IEEEauthorrefmark{1} and Shlomo Shamai (Shitz)\IEEEauthorrefmark{2} \\}

\IEEEauthorblockA{%
\IEEEauthorrefmark{1}
National Chiao-Tung University, Hsinchu, Taiwan\\
E-mail: \texttt{stefano@nctu.edu.tw} }

\IEEEauthorblockA{%
\IEEEauthorrefmark{2}
Technion-Israel Institute of Technology,  Haifa, Israel \\
E-mail: \texttt{sshlomo@ee.technion.ac.il}
}
\thanks{
The work of S. Rini was funded by the  Ministry Of Science and Technology (MOST) under the grant 103-2218-E-009-014-MY2.
The work of S. Shamai was supported by the Israel Science Foundation (ISF).
}
}
\maketitle

\begin{abstract}
The ``writing dirty paper'' capacity result crucially dependents on the perfect channel knowledge at the transmitter as
the presence of even a small uncertainty in the channel realization gravely hampers the ability of the transmitter to pre-code its transmission against the channel state.
This is particularly disappointing as it implies that interference pre-coding in practical systems is effective only when the channel estimates at
    the users have very high precision, a condition which is generally unattainable in wireless environments.
In this paper we show that substantial improvements are possible when the state sequence is drawn from a discrete distribution, such as a constrained input constellation,
for which state decoding can be approximatively optimal.
We consider the ``writing on dirty paper'' channel in which the state sequence is multiplied by a fast fading process
and derive conditions on the fading and state distributions for which state decoding closely approaches capacity.
%
%
%
%
%
%
These conditions intuitively relate to the ability of the receiver to correctly identify both the input and the state realization despite of the uncertainty introduced by fading.
\end{abstract}


\section*{Introduction}

Although interference pre-cancellation is well understood in information theoretical settings, practical implementations of this coding strategy have yet
to find widespread adoption in practical communication systems.
Currently, interference pre-coding can be found only in a few communication standards, usually in its incarnation as Tomlinson-Harashima pre-coding
\cite{harashima1972matched,tomlinson1971new}.
The performance of this implementation is rather low, as compared to the very elegant solution using LDPC codes  and  or trellis-coded quantization \cite{sun2009near}.
One is then bound to wonder as of why high-performing interference pre-cancellation strategies have yet to have a significant impact on communication  systems.
The the answer to this question possibly lays in the intrinsic fragility of this coding technique  which relies
on the specific way in which the desired signal combines with the interference and is thus heavily affected by channel uncertainty.
Many communication systems, instead, utilize interference decoding, which is intuitively a more robust interference management strategy when lacking adequate channel knowledge.
This technique also takes full advantage of the inherent structure of the interference signal, which is often drawn from a finite-rate codebook, as investigated in \cite{simeone2010exploiting},
and/or transmitted using a fixed constellation.
In this correspondence we focus on the latter case and derive    the conditions under which interference decoding is provably close to optimal in the presence of fading and
 partial channel knowledge.

\noindent
{\bf Literature Review:}
The Gel'fand-Pinsker (GP) channel \cite{GelfandPinskerClassic} is a very comprehensive model which, generally speaking, can accommodate for variations
of the ``Writing on Dirty Paper'' (WDP)  channel  to include channel uncertainty and partial side-information.
%
Unfortunately the capacity of the GP channel is expressed as non-convex maximization and a closed-form expression of capacity is available only
 for a handful of models.
%
For this reason, determining the capacity of variations of  Costa's original setup is a challenging task.
In \cite{zhang2007writing}, the authors study the WDP in which the input and the state sequences are multiplied by the same fading coefficient.
%
Here in is shown that the rate loss from full state pre-cancellation is vanishing, since state and input still combine in a predictable manner.
In \cite{RiniPhase14}, we derived the approximate capacity for the WDP channel in which the state is multiplied by uniform binomial fading
by further develop  bounding techniques originally developed in \cite{LapidothCarbonCopying}.
The results in \cite{RiniPhase14} are further extended in \cite{rini2014capacity} to include  more general  fading distribution,
 although restricted to the case of discrete support.

\noindent
{\bf Contributions:}
We investigate the capacity of the ``Writing of Fast Fading Dirt'' (WFFD) channel, a variation of the WDP channel in which the state sequence is multiplied by a fast fading process.
The state realization is assumed to be drawn from a discrete set of values and thus the receiver can attempt to decode both the state and the input realization.
For both models we derive conditions on the support of the fading and state distribution for which state decoding is approximatively optimal.
%
%
The main contribution is the development of an outer bound  which provides sufficient conditions under which state decoding is close to optimal.
We consider  both  the case of \emph{No Channel Side Information} (NCSI) in which fading is not know at either the transmitter or the receiver and
the case of \emph{Receiver Channel Side Information} (RCSI) in which fading is known an the receiver only.
%
%
%

\noindent
{\bf Organization:}
The remainder of the  paper is organized as follows: Sec.  \ref{sec:Channel Model} introduces the channel model while  Sec. \ref{sec:Related Results} presents
relevant results available in the literature.
Sec. \ref{sec:NCSI discrete State} considers the case of no fading knowledge at either the transmitter or the receiver while Sec. \ref{sec:RCSI discrete State}
focuses on the case in which  only the receiver has knowledge of the fading realization.
%
Finally, Sec. \ref{sec:conclusion} concludes the paper.

\noindent
\underline{
Only sketches of the proofs are provided in the main text:}
\underline{the full proofs can be found in appendix.}
\section{Channel Model}
\label{sec:Channel Model}
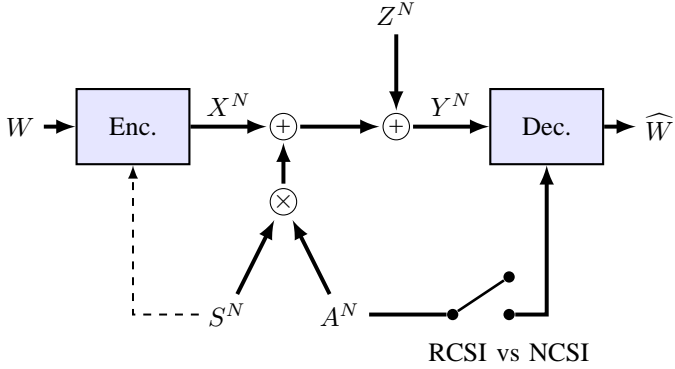
\begin{figure}
\centering
\begin{tikzpicture}[node distance=2.5cm,auto,>=latex]
  \node at (0,0) (source) {$W$};
  \node [int] (enc) [right of = source, node distance = 1.5 cm]{Enc.};
   \node (Pyx) [joint, right of = enc, node distance = 2 cm]{};
   \node (Pyx) [right of = enc, node distance = 2 cm]{+};
   \node (Pyx2) [joint, right of = Pyx, node distance = 1.5 cm]{};
   \node (Pyx2) [right of = Pyx, node distance = 1.5 cm]{+};
    \node [int] (dec) [right of = Pyx2, node distance = 2    cm]{Dec.};
    \node  (dest) [right of=dec, node distance = 1.5 cm] {$\Wh$};
\node (mul) [joint, below of = Pyx, node distance = 1 cm]{};
\node (mul) [below of = Pyx, node distance = 1 cm]{$\times$};
\node (noise) [above of = Pyx2, node distance = 1.5 cm]{$Z^N$};
\node (a) [below of = mul,  node distance = 1.5 cm]{  };
\node (state) [left of = a,  node distance = .75 cm]{$S^N$};
\node (fading) [right of = a,  node distance = .75 cm]{$A^N$};

   \draw[->,line width=1.5 pt] (source) -- (enc);
   \draw[->,line width=1.5 pt] (dec) -- (dest);
   \draw[->,line width=1.5pt] (enc) -- node[above] {$X^N$}(Pyx);
   \draw[->,line width=1.5pt] (Pyx2) -- node[above] {$Y^N$}(dec);
   \draw[->,line width=1.5pt] (Pyx) -- (Pyx2);
   \draw[->,line width=1.5pt] (mul) -- (Pyx);
   \draw[->,line width=1.5pt] (state) --  (mul);
   \draw[->,line width=1.5pt] (fading) --  (mul);
   \draw[->,line width=1.5pt] (noise) -- (Pyx2);
   \draw[->,line width=.75pt,dashed] (state) -|(enc);
 \node (1) [fill,circle,inner sep=0pt,minimum size=0.15 cm , right of = fading, node distance = 1.5 cm] {};
\node (2) [fill,circle,inner sep=0pt,minimum size=0.15 cm,   right of = 1, node distance = .75 cm] {};
\node (3) [fill,circle,inner sep=0pt,minimum size=0.15 cm,  above of = 2, node distance = .5 cm] {};
\draw[-,line width=1 pt] (1) -- (3);
\draw[-,line width=1.5pt] (fading) -- (1);
\draw[->,line width=1.5pt] (2) -| (dec) ;
\node (l2) [below of = 2, node distance = .5 cm]{RCSI vs NCSI};
 \end{tikzpicture}
\caption{``Writing on Fast Fading Dirt with No Channel Side-Information'' (WFFD-NCSI) and the ``Writing on Fast Fading Dirt with Receiver Channel Side-Information'' (WFFD-RCSI).}
\label{fig:WFFD}
\vspace{-.5 cm }
\end{figure}
The ``Writing on Writing on Fast Fading Dirt'' (WFFD) channel is defined as the channel in which the output is obtained as
\ea{
Y^N=X^N+c A^N S^N+Z^N,
\label{eq:fading Dirt Paper Channel general}
}
where the Random Variables (RV) $S^N,A^N$  and $Z^N$ are obtained through iid draws from the distribution $P_S,P_A$ and  $\Ncal(0,1)$ and support $\Scal,\Acal$ and $\Rbb$ respectively.
The sequence $S^N$ is provided non-causally to the transmitter and the channel input $X^N$ is  subject to the constraint $\sum_i^N \Ebb[X_i^2] \leq NP$.
Without loss of generality we assume that $\var[A]=\var[S]=1$ and $\mu_S=0$ so that the variance of the fading-times-state term $c A_i S_i$ is $c^2 \mu_A^2$.
%
%

We further classify the WDP channel in \eqref{eq:fading Dirt Paper Channel general} with respect to the available channel side-information:

\noindent
$\bullet${\bf WFFD with No Channel Side-Information (WFFD-NCSI):} the fading sequence $A^N$ is not know at either the transmitter or the receiver.

\noindent
$\bullet${\bf  WFFD with Receiver Channel Side-Information (WFFD-RCSI):} the fading sequence $A^N$ in know at only at the receiver.

The  WFFD-RCSI is obtained from the WFFD-NCSI by providing the sequence $A^N$ as an additional channel output, that is
\ea{
Y_{\rm RCSI}^N =[ Y_{\rm NCSI}^N \ A^N ],
\label{eq:channel output RCSI}
}
for $Y$ in \eqref{eq:fading Dirt Paper Channel general}.
%
A  graphical representation of these two channel models is provided in Fig. \ref{fig:WFFD}:
the switch on the noiseless channel between $A^N$ and the receiver indicates whether the fading side-information is available to the receiver or not.
The dotted line between $S^N$ and the transmitter represents the anti-causal channel knowledge at the transmitter.
Standard definitions of rate, code, achievable rate, capacity and approximate capacity are assumed.
%

In the following we consider the case in which $\Scal$ is a discrete set: a recurring example is the case in which  $S^N$ is
uniformly distributed over the PAM input constellation
\ea{
\Scal_{m-PAM}= \lcb \p{
 2 i \Delta_m,    \   i \in  \lsb \f {1-m} 2 \ldots  \f {m-1} 2 \rsb  & m \rm  \ even \\
 (2 i + 1) \Delta_m, \ i \in  \lsb -\f m 2 \ldots  \f m 2-1  \rsb           & m \rm  \ odd
} \rnone
\label{eq:pam support}
}
for $\Delta_m=\sqrt{3 / (m^2-1)}$
which guarantees $\var[S]=1$ and $\mu_S=0$ as by assumption.
%

\section{Related Results}
\label{sec:Related Results}
\medskip
\noindent
$\bullet$ {\bf ``Gelfand-Pinsker'' (GP) channel:}
The capacity of the  GP channel \cite{GelfandPinskerClassic} is a classic result and is expressed as
\ea{
\Ccal=\max_{P_{U,X|S}} \lb  I(Y; U) - I(U;S) \rb.
\label{eq:Capacity of GP channel}
}
The capacity of both the WFFD-NCSI and the WFFD-RCSI can be evaluated through \eqref{eq:Capacity of GP channel}.
Unfortunately the expression in \eqref{eq:Capacity of GP channel} is convex in $P_{X|S,U}$ for a fixed $P_{U| S}$
but neither convex nor concave in $P_{U|S}$ for a fixed $P_{X|S, U}$: consequently this expression
cannot be easily obtained in a closed-form or numerically approximated.

\begin{figure}
\begin{center}
\begin{tikzpicture}
\node at (-4.5,0) {\includegraphics[trim=0cm 0cm 0cm 0cm,  ,clip=true,scale=0.42]{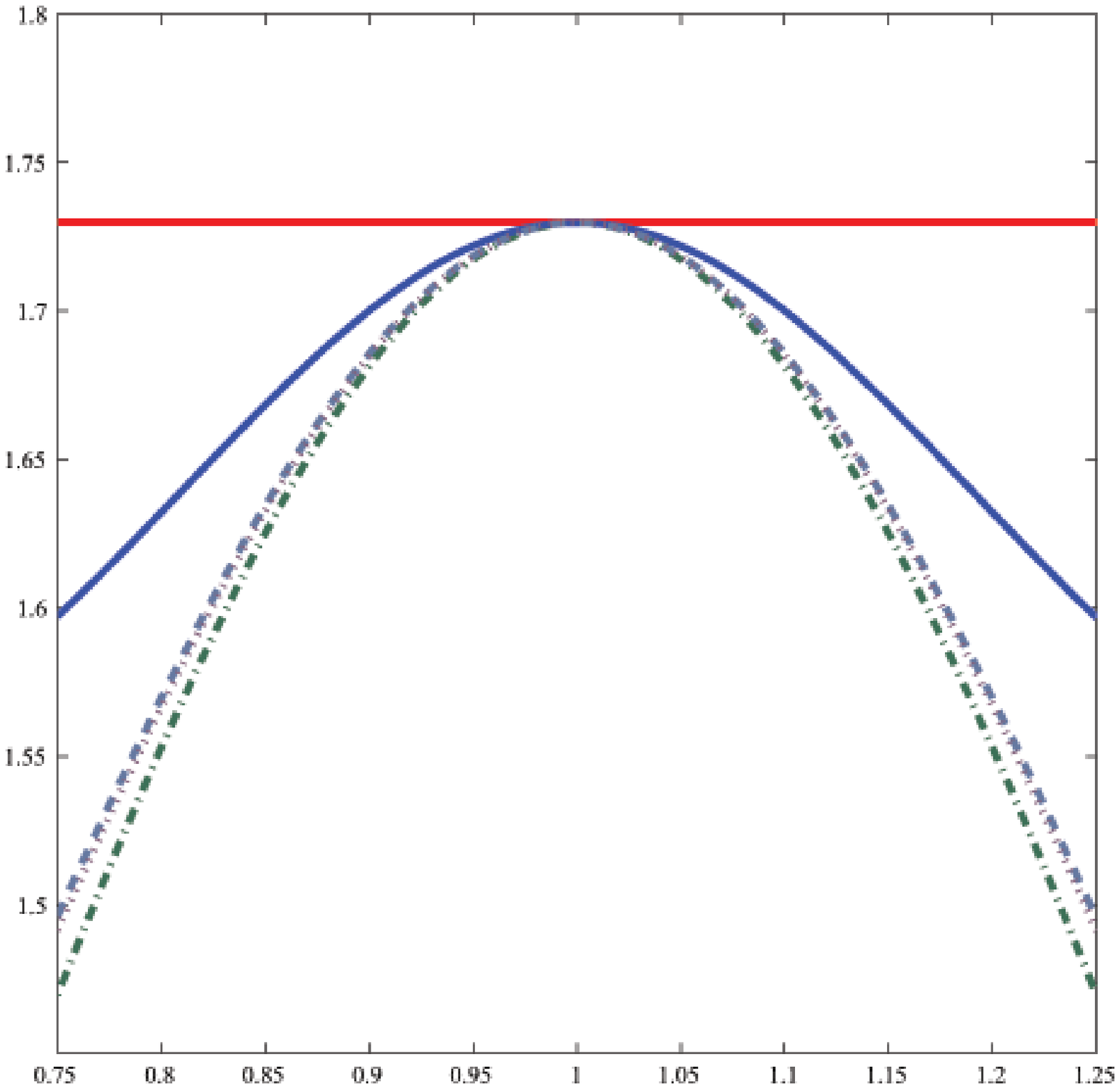}};
\node[rotate=90] at (-8.5,0) {{$R~[bits]$ }} ;
\node at (+.34-4.5,-3.8) {$k$};
\draw[dashed] (+.34-4.5,-3.2) -- (+.34-4.5,3.5);
\node[rotate = 0] at (-6,2.5) {\color{red} $\Ccal$};
\node[rotate = 0] at (-6.5,1.25) {\color{blue} 2-PAM};
\node[rotate = -0] at (-6.5,-2.5) {\textcolor[rgb]{0.00,0.50,0.00}{$\Ncal$ }};
\node[rotate = -0] at (-6,-1.5) {\textcolor[rgb]{0.50,0.00,0.50}{6-PAM}};
\node[rotate = -0] at (-5.5,-0.5) {\textcolor[rgb]{0.50,0.50,1.00}{4-PAM}};
\vspace{-.7 cm}
\end{tikzpicture}
\vspace{-.3 cm}
\caption{The mismatch loss for $P=10$ and $c=5$ and when $S$ is an equiprobable PAM signal (2,4 and 6-PAM) or a Gaussian sequence ($\Ncal$).}
\label{fig:mismatched}
\end{center}
\vspace{-.9 cm}
\end{figure}

\medskip
\noindent
$\bullet$ {\bf ``Writing on Dirty Paper'' (WDP) channel:}
Consider a WDP channel and assume that, given the imperfect channel knowledge at the transmitter, the encoder believes the state sequence to be $c k S^N$ instead of $c S^N$.
The rate loss due to the imperfect channel estimation can be readily evaluated as in Fig. \ref{fig:mismatched}.
The largest rate loss corresponds to the case in which $S$ is Gaussian distributed, in which case
\ea{
\Ccal-R^{\rm DPC \ \Ncal}(k)=\f 1 2 \log \lb 1+ \f {P a^2 }{P+a^2+1} (k-1)^2 \rb,
\label{eq:gaussian mismatch loss}
}
and is obtained from the expression in \eqref{eq:Capacity of GP channel} by letting $U=[X \ S]$
%
while the rate loss is increasing with $m$ when $S$ is an equiprobable m-PAM.

\medskip
\noindent
$\bullet$ {\bf GP channel with state amplification:}
%
%
The GP channel in the case in which the transmitter is required to decode both the transmitted message and the channel state is known as GP with ``state amplification''
\cite{kim2008state}.
The largest transmission rate $R$ that can be attained in this channel
\ea{
R^{\rm IN-SA} = \max_{P_{X|S}} I(Y;X,S)-H(S),
\label{eq:full state amplification}
}
and corresponds to the expression in \eqref{eq:Capacity of GP channel} for the choice $U=[X \ S]$, that is the decoded message corresponds to both the channel input and the state.
This is the attainable rate in a point-to-point channel in which two codewords, $X^N$ at rate $R$ and $S^N$ at rate $H(S)$, are transmitted over the channel.

\begin{figure}
\begin{center}
\begin{tikzpicture}
\node at (-6,0) {\includegraphics[trim=0cm 0cm 0cm 0cm,  ,clip=true,scale=0.43]{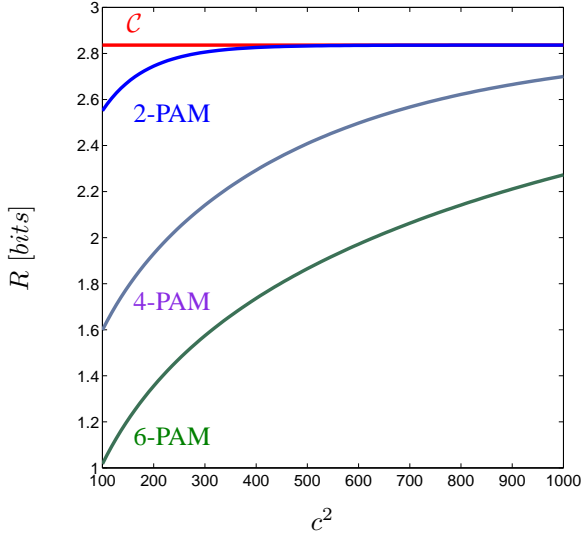}};
\node[rotate=90] at (-8-2,0) {{$R~[bits]$ }} ;
\node at (0-6,-3.6) {$c^2$};
\node[rotate = 0] at (0-8.5,3) {\color{red} $\Ccal$};
\node[rotate = 0] at (-2-6,1.8  ) {\color{blue} 2-PAM};
\node[rotate = -0] at (-4-4,-.7) {\textcolor[rgb]{0.54,0.17,0.89}{4-PAM}};
\node[rotate = -0] at (-4-4,-2.5) {\textcolor[rgb]{0.00,0.50,0.00}{6-PAM}};
\vspace{-.5 cm}
\end{tikzpicture}
\caption{The state amplification performance for transmit $P=100$ and state power $c^2 \in [10^2 \ldots 10^3]$ for the case in which the state
has is PAM constellation (2,4 and 6-PAM).  }
\label{fig:StateAmp}
\end{center}
\end{figure}

\section{WFFD-NCSI Channel}
\label{sec:NCSI discrete State}
\begin{figure}
\begin{center}
\begin{tikzpicture}
\node at (-6,0) {\includegraphics[trim=0cm 0cm 0cm 0cm,  ,clip=true,scale=0.42]{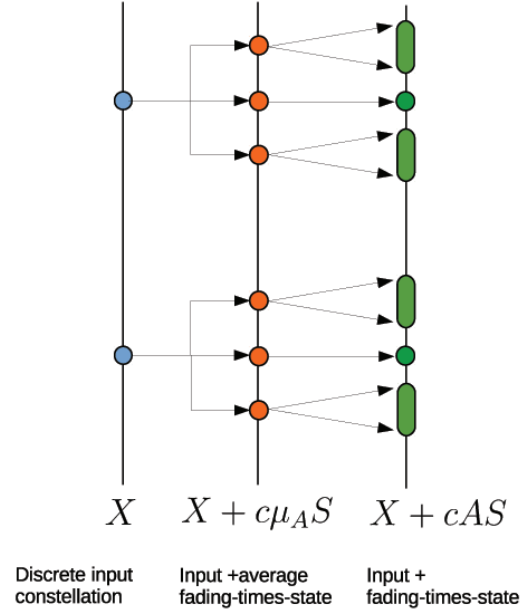}};
%
\end{tikzpicture}
\vspace{-.5 cm}
\caption{A representation of the output space in the WRDP-NCSI channel with discrete state.}
\label{fig:AlignedSet3}
\end{center}
\end{figure}

In the WFFD-NCSI neither the transmitter nor the receiver have knowledge of the exact way in which the channel input collides with the fading-times-state term $cA^NS^N$.
For this reason, pre-coding as in the WDP channel is effective only when the overall variance of the term $c A^NS^N$ is small, in which case the
users still incur in a loss similar to the one in \eqref{eq:gaussian mismatch loss}.
An alternative strategy is for the receiver to decode both the state realization, along with the transmitted message.
To facilitate this, the transmitter can restrict its input to a finite constellation such that the receiver can decode both $X^N$ and $S^N$ from the channel output, as shown in Fig. \ref{fig:AlignedSet3}.
This figure conceptually represent how the fading affects the channel output: the random effect of $A$ is to ``spread'' the value $c AS$
in an interval around the values $c \mu_AS$.
By restricting the channel input to be discrete and sufficiently spaced apart, the support of $X+c AS$ is composed of
non-overlapping sets for different $X$ and $S$, regardless of the realization of $A$.
When the minimum distance between these sets is sufficiently large, the receiver can decode both $X$ and $S$ from $Y$ with high probability.
This intuition is formalized in the next theorem.

\begin{thm}{\bf Outer bound and approximate capacity for the WFFD-NCSI with discrete state. \\}
\label{th:NCSI out}
Consider the WFFD-NCSI in Fig. \ref{fig:WFFD} with $P,c^2 >1$  and for  $\Scal$ and $\Acal$
such that
\ea{
\min_{
\small \p{s,\st \in \Scal, s>\st, \
 a,\at \in \Acal   \\
i \in \lsb -2\lceil \sqrt{P} \rceil \ldots +2\lceil \sqrt{P} \rceil \rsb}} |i -c a s- \at \st|> \f 12,
\label{eq:conditions NCSI out}
}
then capacity $\Ccal$ is upper bounded as
\ea{
\Ccal \leq R^{\rm OUT} = \max_{P_{X|S}}I(Y;X,S) - H(S)+4, 
\label{eq:NCSI OUT}
}
and the exact capacity is to within $15 \ \bpcu$ from the outer bound in \eqref{eq:NCSI OUT}.
\end{thm}
\begin{IEEEproof}
See App. \ref{app:NCSI out}.
\end{IEEEproof}
%
%
The conditions in \eqref{eq:conditions NCSI out} indeed reflect the interpretation in Fig. \ref{fig:AlignedSet3}:
this term is the smallest distance between two contiguous regions in $X+cAS$ when $X$ is restricted to be an integer number in $\lsb \lceil \sqrt{P} \rceil \ldots +\lceil \sqrt{P} \rceil \rsb$.
This follows from the fact that restricting the input to this interval has a small effect on capacity, both from the inner and the outer bound perspective.

The main challenge in proving Th. \ref{th:NCSI out} is in the bounding thorough a closed-form expression of the capacity of WFFD-NCSI as obtained from the capacity of the GP channel in \eqref{eq:Capacity of GP channel}.
Note that
\ea{
I(U;Y) - I(U;S)=I(X,S;Y) - H(S) + H(S|Y,U),
}
so that the state amplification lower bound in \eqref{eq:full state amplification} is close to capacity when $H(S|Y,U)$ is close to zero.
Determining the optimality of state decoding therefore entails  showing that the entropy of $H(S|Y,U)$ is small for the optimal choice of $P_{XU|S}$ in \eqref{eq:Capacity of GP channel}.
To prove this we build upon an outer bounding technique originally introduced in \cite{JafarConjecture} which itself stems from
 the earlier work of \cite{bresler2008two} on the deterministic approximation of AWGN multi-terminal  channels.

As an example of the conditions in \eqref{eq:conditions NCSI out} consider the case in which $S$ is a $m$-PAM sequence (assume $m$ even for convenience) while $A$ has a continuous uniformly distribution:
by restricting the channel input to an integer constellation, the term $X+cAS$ has support $\bigcup_{ij} \Rcal_{ij}$ for
\ea{
\Rcal_{ij} &= [i+ 2j \Delta_m  c (\mu_A -\sqrt{3}),i+2j \Delta_m c\Delta(\mu_A+\sqrt{3})j],
\label{eq:subsets}
}
with $j \in [ -\lfloor \sqrt{P} \rfloor \ldots  \lfloor \sqrt{P}\rfloor ]$  and $i \in \lsb \f {1-m} 2 \ldots  \f {m-1} 2 \rsb$.
Any value of $c$ and $\mu_A$ which guarantees that the above regions are separated of more than one half, satisfies the condition in \eqref{eq:conditions NCSI out}.
A less general result can obtained by requiring the specific order in which the subsets in
\eqref{eq:subsets}.
For instance we could require that $\Rcal_{ij} < \Rcal_{i(j+1)} < \Rcal_{(i+1)1}$ or equivalently $i+cAs \leq i+cA(s+2\Delta_m) \leq (i+1)+cA(1-m) \Delta_m$ for all values of $A$.
For this ordering of the sets $\Rcal_{ij}$, the minimum distance between two contiguous sets $\Rcal_{ij}$ is
\ea{
\min \lcb 2 \Delta_m c (\mu_A -(2i-1) \sqrt{3}), 1-4 \Delta_m c (m-1)\mu_A \rcb,
}
and the result in Th. \ref{th:NCSI out} applies when $D>1/2$.

\section{WFFD-RCSI Channel}
\label{sec:RCSI discrete State}
\begin{figure}
\begin{center}
\begin{tikzpicture}
\node at (-7,0) {\includegraphics[trim=0cm 0cm 0cm 0cm,  ,clip=true,scale=0.42  ]{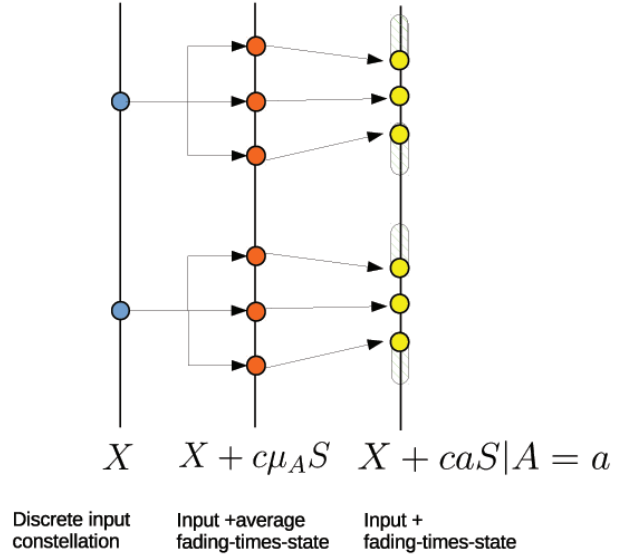}};
%
\end{tikzpicture}
\caption{A representation of the output space in the WRDP-RCSI channel with discrete state.}
\label{fig:AlignedSetRCSI2}
\vspace{-.75 cm}
\end{center}
\end{figure}

In the WFFD-RCSI the receiver fading knowledge reduces the uncertainty on the way in which the input and state combine to produce the channel output.
Unfortunately state pre-coding as in the WDP channel appears to still not be feasible as no known distributed strategy can produce a signal in which the input and the channel state sum in away which is predictable for the transmitter \cite{zamir2012anti}.
On the other hand,  when the state is drawn from a discrete support, state decoding remains a natural transmission strategy to be considered.
In this model, given the additional fading knowledge, the receiver knows which linear combination of the input and state it observes in the channel output, although both the input and the channel states remain unknown.
This is  conceptually represented in presented in Fig. \ref{fig:AlignedSetRCSI2}:  as for the WFFD-NCSI, the effect of fading is to spread the value $c \mu_A S$ in the
interval $c A S$.
While the transmitter has no knowledge of this random effect, the receiver knows which linear combination of input and state is present in the output.
When the input is restricted to a finite constellation, the sum of state and input can be decoded with high probability whenever the support of $X+aS$ is composed of sufficiently separated elements for all possible $a \in \Acal$.
The difference between state decoding in the WFFD-NCSI and WFFD-RCSI can be visualized  by comparing Fig. \ref{fig:AlignedSet3} and Fig. \ref{fig:AlignedSetRCSI2}: %
since the receiver in the WFFD-RCSI has knowledge of $A$, it does not need to account for the ``spread'' of the value $cAS$ but instead has to know that these values are distinguishable for all possible realizations $A=a$.
%
%
%
%
As for Th. \ref{th:NCSI out}, the next theorem formalizes this intuition.

\begin{thm}{\bf Outer bound and approximate capacity for some discrete state distributions.\\}
\label{th:RCSI OUT}
Consider the WFFD-RCSI in Fig. \ref{fig:WFFD} with $P,c^2 >1$  and for  $\Scal$ and $\Acal$
 such that
\ea{
\min_{s,a, i \in [-2\lfloor \sqrt{P}\rfloor ... +2\lfloor \sqrt{P}\rfloor] } |i -c a(s- \st)|>\f 12,
\label{eq:RCSI OUT conditions}
}
then capacity $\Ccal$ is upper bounded as
\ea{
\Ccal \leq R^{\rm OUT} = \max_{P_{X|S}} I(Y;X,S|A)   - H(S)+ 6,
\label{eq:RCSI OUT}
}
and the exact capacity  is to within  $6 \  \bpcu$ from the outer bound in \eqref{eq:RCSI OUT}.
\end{thm}
\begin{IEEEproof}
See App. \ref{app:RCSI, general 2}.
\end{IEEEproof}
Th.  \ref{th:RCSI OUT} is the analog of Th. \ref{th:NCSI out} for the WFFD-RCSI and again the main contribution is the developing an outer bound to the capacity expression in
\eqref{eq:Capacity of GP channel} which matches the state decoding inner bound.
%
%
The difference in the conditions of Th. \ref{th:RCSI OUT}  and  those in Th. \ref{th:NCSI out} also reflects the difference between Fig. \ref{fig:AlignedSetRCSI2} and Fig. \ref{fig:AlignedSet3}: since the receiver knows the realization $A=a$, the elements that must be distinguished are the terms in $X+c aS$ instead of the interval $X+c AS$.
%
%

\begin{figure}
\begin{center}
\begin{tikzpicture}
\node at (-7,0) {\includegraphics[trim=0cm 0cm 0cm 0cm,  ,clip=true,scale=0.46]{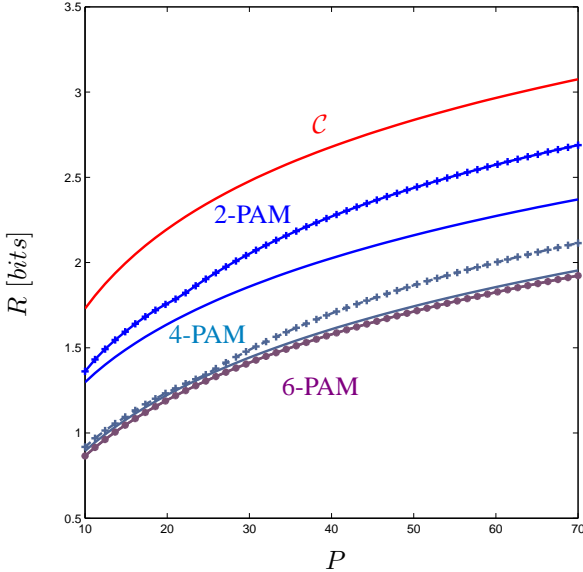}};
\node[rotate=90] at (-11,0) {{$R~[bits]$ }} ;
\node at (-6.8,-3.8) {$P$};
\node[rotate = 0] at (-7,2) {\color{red} $\Ccal$};
\node[rotate = 0] at (-7.9,.8) {\color{blue} 2-PAM };
\node[rotate = -0] at (-8.5,-.8) {\textcolor[rgb]{0.00,0.50,0.75}{4-PAM}};
\node[rotate = -0] at (-7,-1.5) {\textcolor[rgb]{0.50,0.00,0.50}{6-PAM}};
\end{tikzpicture}
\vspace{-.55 cm}
\caption{Attainable rates with Gaussian signaling for the case of no transmitter state information knowledge (plain lines) versus
versus the case of transmitter state knowledge (dotted lines)
 for $c=2$, $A \sim \Ncal(0,1)$ and $P\in [10, 70]$ .}
\label{fig:attainable uniform plus linear}
\end{center}
\vspace{-1 cm}
\end{figure}

It is interesting to compare the performance of the WFFD-RCSI with the performance of the same model but where the transmitter does not have anti-causal knowledge of the
state sequence.
\begin{lem}{\bf Performance without transmitter state knowledge.\\}
\label{lem:Performance without transmitter state knowledge}
If the transmitter does not posses state anti-causal knowledge of $S^N$, then the capacity
of the WFFD-RCSI can be outer bounded as
\ea{
\Ccal
&= \max_{P_X} \ I(Y;X|A).
\label{eq:Performance without transmitter state knowledge}
}
\end{lem}
The result in Lem. \ref{lem:Performance without transmitter state knowledge} follows naturally from the point-to-point capacity result.
%
The RHS of \eqref{eq:RCSI OUT conditions} can be rewritten as
\ea{
I(Y;X,S|A) - H(S) = I(Y;X|A) - H(S|X,A,Y),
\label{eq:no state 2}
}
and, by comparing  \eqref{eq:no state 2} to \eqref{eq:RCSI OUT},  one would be tempted to conclude that channel knowledge does not provide much rate advantages.
%
It must be noted  that  the maximization in \eqref{eq:RCSI OUT}  and \eqref{eq:Performance without transmitter state knowledge} are performed over two different set of distributions: the first maximization is over $P_{X|S}$ while the latter is over $P_{X}$.
In general, it is not easy to determine the rate improvement provided by this enlarged optimization set, especially because linear strategies are usually not optimal.
To illustrate this point, we can again return to the example where $S$ is a 2-PAM sequence while $A$ is uniformly distributed with mean $\mu_A$:
when state knowledge is available at the transmitter, it can use part of its power to remove the effect of the mean of the fading realization by choosing
\eas{
\Xt & \sim \Ncal(0,1) \\
X & =\al \Xt -  \sqrt{1-\al^2} K,
\label{eq:linear signaling}
}
for some RV $K$  with zero mean and unit variance so that
\ea{
Y|A=\al X+SA +K \sqrt{1-\al^2} + Z.
}
For example the choice
\ea{
K=\lcb\p{
-  1 & S=+1 \\
+ 1  & S=-1
}
\rnone
}
can be used to increase the entropy of $Y|A$ when $S$ is a 2-PAM sequence.
%
%
The rate improvements provided by this strategy with respect to the case of no channel state information and Gaussian signaling are presented in Fig. \ref{fig:attainable uniform plus linear} for the case in which the state sequence is an equiprobable  m-PAM sequence.

%
%

\section{Conclusions}
\label{sec:conclusion}

In this paper we have identified cases where interference decoding
aided by an interference cognitive transmitter, which happens to be
the more common practice, is close to capacity in a number of scenarios
which also include fading.
More specifically, we study the capacity of the ``writing on fast fading dirt'' channel,  a variation of the classical ``writing on dirty paper'' channel in which the channel state is multiplied by a fast fading sequence.
The channel state  il also assumed to have a discrete support, modelling an interference signal from a constrained constellation
We consider two scenarios: (i) the case in which neither the transmitter nor receiver have side-information and (ii) the case in which only the receiver has knowledge of the fading process.
In both cases we derive conditions on the support of the fading and state distribution so that state decoding is to within few bits from capacity.
These conditions intuitively relate to the ability of the decoder to distinguish both the channel input and the state realization from the channel output, regardless of the noise realization.
These models are a special case of the Gelfand-Pinsker channel for which capacity is known but expressed as the solution  of a non-convex optimization problem.
For this reason, our approximate capacity result entails a careful bounding of the capacity expression to yield a closed-form outer bound.
%
%
%
%
%

%
\bibliographystyle{IEEEtran}
\bibliography{steBib3,steBib1}

\newpage
\onecolumn

\section{Proof of Th. \ref{th:NCSI out}.}
\label{app:NCSI out}

The proof is shown by proving the outer bound in \eqref{eq:NCSI OUT}, since the achievability follows trivially from \eqref{eq:full state amplification}.
%
The this outer bound follows the derivation in \cite{JafarConjecture} in which a similar procedure is employed to investigate the degrees of freedom of the broadcast channel with finite precision CSIT.
%
%
The main difference is in that we retain the additive noise in the channel with integer, peak-limited input: this makes it possible to express the outer bound as a maximization over the same class of distribution as the inner bound in \eqref{eq:full state amplification}.
%

\begin{figure}
\centering
\begin{tikzpicture}
    \node at (0,0) [joint] (c1) {};
    \node at (2,0) [joint] (c2) {};
    \node at (-2,0) [joint](c3) {};
    \node at (4,0) [joint] (c4) {};
    \node [below=of c1.west,node distance=.25 cm, anchor=west,rotate = -45] (l1) {$\Ccal$ original capacity};
    \node [below=of c2.west,node distance=.25 cm, anchor=west,rotate = -45] (l2) {$\overline{\Ccal}$ integer, peak limited input};
    \node [below=of c3.west,node distance=.25 cm, anchor=west,rotate = -45] (l3) {$R^{\rm IN-SA}$  state amplification \eqref{eq:full state amplification}};
    \node [below=of c4.west,node distance=.25 cm, anchor=west,rotate = -45] (l4) {$R^{\rm OUT}$ using \cite{JafarConjecture}};
\draw (c1) -- (c2);
\draw (c1) -- (c3);
\draw (c2) -- (c4);
 %
\end{tikzpicture}
\caption{A conceptual representation of the proof in Th. \ref{th:NCSI out}.}
\label{fig: bounds plot}
\end{figure}
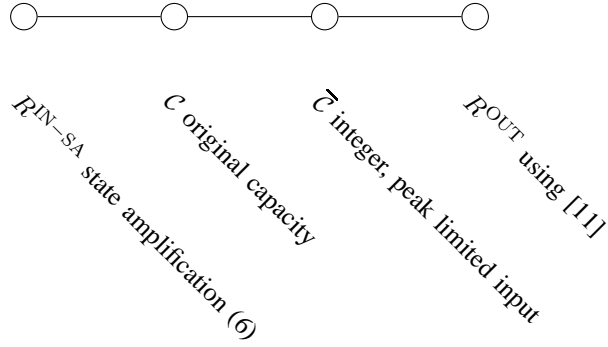

These steps are conceptually presented in Fig. \ref{fig: bounds plot}, where

\noindent
$\bullet$ $\Ccal$ is the actual capacity,  which can be obtained through the result in \eqref{eq:Capacity of GP channel}, this value is to within small gap from $\overline{\overline{\Ccal}}$.

\noindent
$\bullet$ $\overline{\overline{\Ccal}}$ is the capacity of the channel in which the input is restricted to integer, peak limited values. The capacity of this channel is also obtained through the result in \eqref{eq:Capacity of GP channel} but it can be further upper bounded through $R^{\rm OUT}$.

\noindent
$\bullet$ $R^{\rm OUT}$ is the upper bound in $\overline{\overline{\Ccal}}$ obtained through the ``aligned message set'' approach in \cite{JafarConjecture} and which is expressed only
as maximization over all the distributions $P_{X|S}$.

\noindent
$\bullet$ $R^{\rm IN-SA}$ is the state amplification inner bound in \eqref{eq:full state amplification} and it has the same expression as $R^{\rm OUT}$ but for an additive term.

%

\bigskip
\noindent
$\bullet$ { \bf  Integer, peak-limited channel:}
\medskip
The first step in the proof is to show that the capacity of the noiseless channel in which the inputs are restricted to be integers and peak-limited at $\sqrt{P}$
 is close in capacity to the channel of the original channel.
Let
\eas{
\Yo^N &=\lfloor X^N \rfloor + c A^N S^N+\Zo^N \\
  E^N &= Y^N-\Yo^N = X^N-\lfloor X^N \rfloor-Z^N+\Zo^N,
\label{eq: integer channel}
}
where $\Zo^N$ has the same distribution of $Z^N$ but is independent from it.
that is,  $\Yo^N$ is the WFFD-NCSI in which the input is restricted to be integer-valued and power constrained;
we then have
\eas{
N(R-\ep)
& \leq  I(Y^N;W)\\
& \leq  I(Y^N,E^N;W) \\
& \leq I(\Yo^N,E^N;W)
\label{eq:p1 1}\\
& \leq I(\Yo^N;W)+I(E^N; W | \Yo^N )\\
& \leq I(\Yo^N;W)+H(E^N)-H(E^N| W , X^N, \Yo^N )\\
& \leq I(\Yo^N;W)+\f N 2 \log(2 \pi e (\var(X^N-\lfloor X^N \rfloor)+2)-H(\Zo^N-Z^N|Z^N)
\label{eq:p1 2}\\
& \leq I(\Yo^N;W)  + \f N 2 \log(2 \pi e 3)- \f N 2 \log \lb 2 \pi e \rb \\
& \leq I(\Yo^N;W)+\f N 2 \log 3
}{\label{eq:p1}}
where  \eqref{eq:p1 1}  follows from the fact that the transformation of variables has unitary Jacobian and \eqref{eq:p1 2} follows from the fact that the variance of a random variable bounded in $[a \ b]$  is upper bounded by the variance of discrete random variable that takes values $a$ and
$b$ with equal probability.

The inequality in \label{eq:p1} establishes that the capacity of the integer-valued channel is at most $0.8 \ \bpcu $ larger than the capacity of the original channel.
Note that $\Yo^N$ is equal to $Y^N$ but for the additive noise but $Z^N$ is replaced with the identical, independent noise $\Zo^N$.

We now wish to further restrict the channel to have a peak power constraint instead of an average power constraint.
To do so we define
\eas{
\Xoo^N & = \lfloor X^N \rfloor  \mod \lceil \sqrt{P} \rceil, \\
\Xt^N  &=  \lfloor X^N \rfloor   - \Xoo^N, \\
\Yoo^N & = \Xoo^N +A^N S^N+\Zo^N,
}{\label{eq:peak constraint output}}
and once again we use Fano's inequality to write
\eas{
I(\Yo^N;W)
& \leq I(\Yo^N,\Xt ;W) \\
& \leq I(\Yoo^N, \Xt ;W)
\label{eq:p2 1} \\
& \leq I(\Yoo^N ;W) + I(\Xt^N; W | \Yoo^N) \\
& \leq I(\Yoo^N ;W) + H(\Xt^N)-H(\Xt^N| W , \Yoo^N)\\
& \leq I(\Yoo^N ;W) + H(\Xt^N) \\
& \leq I(\Yoo^N ;W) + N \max_j H(\Xt_j),
\label{eq:p2 2}
}{\label{eq:p2}}
where \label{eq:p2 1} follows from the fact that this transformation has unitary Jacobian and \label{eq:p2 2} from the fact that $\Xt$ in a discrete random variable with positive
defined entropy.
We are now left with the task of bounding the term  $H(\Xt_j)$ which can be done as in \cite[(156)-(158)]{JafarConjecture}.
Using the bound in \cite[(156)-(158)]{JafarConjecture} and 
 in \eqref{eq:p2}  we can conclude that the capacity of WFFD-NCSI where the inputs are integer and peak-limited is to within a
constant gap from the capacity of the general WFFD-NCSI.
This is because the proof in \cite{GelfandPinskerClassic} is developed from Fano's inequality which is tight in this model.
%
%
%
Next we derive an upper bound to the capacity of the WFFD-NCSI with integer, peak-limited channel inputs.

\medskip
\noindent
$\bullet$ {\bf  Capacity outer bound:}
\medskip
The capacity of the WFFD-NCSI is determined by the result in \eqref{eq:Capacity of GP channel}:
this expression can be further manipulated as
\ea{
I(\Yoo;U)-I(S;U) = I(\Yoo;S,\Xoo) - H(S)+H(S|\Yoo, U),
}
where we have used the fact that  $\Xoo$ can be taken to be a deterministic function of $S$ and $U$ and the Markov chain $U-S \ \Xoo-\Yoo$.
Additionally the term $H(S|\Yoo, U)$ can be rewritten as
\eas{
& H(S|U, \Yoo )  \\
& \leq H(S, [Z]|U, \Yoo )  \\
& = H([Z]) + H (S |U, X+AS+Z-[Z] ).
\label{eq:p5 2}
}
Let's now bound $H(S|U, \Yoo)$ as:
\eas{
& H(S|U, \Yoo )  \\
& \leq H(S, [Z]|U, \Yoo )  \\
& = H([2Z]/2) + H (S |U, X+AS+Z-[2Z]/2 ) \\
& = H([2Z]/2) + H (S |U, X+AS+\Zh )\\
& = H([2Z]/2) + H (S |U, \Yh ),
\label{eq:p5 2}
}
where $[Z]$ indicates the integer part of $Z$, that is
\ea{
[Z]=\lcb  \p{
\lfloor Z \rfloor & Z \geq 0 \\
\lceil Z \rceil & Z < 0
}\rnone
}
while $\Zh=Z-[2Z]/2$ is  noise bounded in the interval $[-1/4,+1/4]$ and $\Yh$ is the output corresponding to the channel where the channel noise is $\Zh$.
The RV $Z-\Zh=[2Z]/2$ is a discrete random variable with a finite positive entropy which we can bounded as
\eas{
H ([2Z]/2)
& =  \Pr \lsb Z \in \lb  - \f 1 4, + \f 1 4\rb \rsb - \sum_{i \in \Nbb} 2 \rho_z \log (\rho_z).
\label{eq:rho z}
}
for
\ea{
\rho_z=P\lsb Z \in \lb \f i 2- \f 1 4, \f i 2+ \f 1 4\rb\rsb.
}
For $|i|>1$ we have
\ea{
\rho_z < 0.1747  \leq e^{-1} \approx 0.3679
}
and therefore the terms  $-\rho_z \log \rho_z$  in the RHS of \eqref{eq:rho z} are decreasing in $\rho_z$; consequently we can use usual inner and lower on the $Q$ function to
write:
\ean{
\rho_z
& \leq \f 12 e^{-\f {(i/2-1/4)^2} 2 }-\f 1 { \sqrt{2\pi} (i+1/2) } \lb 1 - \f 1 {(i/2+1/4)^2}\rb  e^{-\f {(i/2+1/4)^2} 2 } \\
& = e^{-\f  {(i-1/2)^2} 8 }  \lb \f 1 2  - \f 1 {\sqrt{2\pi} (i+1/2) } \lb 1 - \f 1 {(i+1/2)^2}\rb e^{- \f i 4} \rb \\
& \leq \f 12 e^{-\f  {(i-1/2)^2} 8 }.
}
The function $\exp\{-\f  {(i-1/2)^2} 8 \}$ is monotonically decreasing for $i\geq 4$, so that
\ean{
& \sum_{i=4}^{\infty}  -\rho_z \log \rho_z   \\
& =\sum_{i=4}^{\infty}   \f  {(i-1/2)^2} 8 e^{-\f  {(i-1/2)^2} 8 } \\
& \geq  \int_{i=4}^{\infty}  \f  {(i-3/2)^2} 8 e^{-\f  {(i-3/2)^2} 8 } \\
& =  1.21,
}
so that now we can write
\ea{
H ([Z]) & \leq  0.54 + \sum_{i=1}^3 -\rho_z \log \rho_z + 1.21 \nonumber \\
& =   0.54 + 2.14+ 1.21 \leq 4.
%
\label{eq:p6}
}
Using \eqref{eq:p6}, we can further bound  \eqref{eq:p5 2} as
\ean{
H(S|U, \Yoo )
& = H (S |U, \Yh) + 4\\
& \leq H (S | \Yh) + 4 \\
& \leq \log (Q_S (\Yh))+4.
}
%
%
%
And where $Q_S (\Yh)$ is the set of $\st \in \Scal$ for which there exist $\at \in \Acal$  and $\ut \in \Ucal$ and $\zh \in [-1/4,+1/4]$ such that
\eas{
& \Xoo(\st,\ut)+c \at \st +\zt= \yh,
}
that is, it is the set of all possible $S=\st$ that could have produced the output $\Yh=\yh$.
We next want to find the conditions under which the cardinality of $Q_S (\Yh$ is always one.
This can be done assured when the images of the output under a noise bounded by between $0$ and $1$ which is granted when
\ea{
\min_{u,\st,s,\ut} |\Xoo(s,u)+c s a -(\Xoo(\st,\ut)+c \at \st)|> \f 1 2,
}
since $\Xoo$ only takes values over the integers, we have
\ea{
\min_{\st,s,i \in -\sqrt{P} ... +\sqrt{P}} |i+c (s a -\at \st)|> \f 1 2 ,
}

Finally we obtain that, when condition \eqref{eq:conditions NCSI out} we have that $\Ccal$ is to within $14 \ \bpcu$ from the outer bound
\ean{
& \max_{ P_{\Xoo|S}} I(\Yoo;U)-I(S;U)  \\
& \leq \max_{ P_{\Xoo|S}} I(\Yoo;S, \Xoo)-H(S)+4.
}
On  the other hand, by enlarging the class of input distribution for the channel, we have
\ea{
R^{\rm OUT} = \max_{ P_{X|S}} I(\Yoo;S, \Xoo)-H(S)+4,
}
which corresponds to the outer bound is \eqref{eq:NCSI OUT}.
%
%
%
%
%

\section{Proof of Th. \ref{th:RCSI OUT}.}
\label{app:RCSI, general 2}
The state amplification inner bound in  \eqref{eq:full state amplification}  for the channel output in \eqref{eq:channel output RCSI} yields the attainable rate
\eas{
R^{\rm IN-SA} & = \max_{P_{X|S}} I(Y;X, S|A)-H(S)
\label{eq:p5 V2}
}
and thus, as for the proof of Th. \ref{th:NCSI out}, the theorem is shown by deriving the outer bound in \eqref{eq:RCSI OUT}.

%
%
%
As for the proof of Th. \ref{th:NCSI out} in App. \ref{app:NCSI out}, we rely on the fact that the capacity of the WFFD-RCSI to a deterministic, integer, peak-limited channel.
The derivation is substantially the same as in App. \ref{app:NCSI out}, since the WFFD-RCSI differs from the WFFD-NCSI in that it has $A$ as an extra output.
It can be verified that this difference does not affect the derivation in App. \ref{app:NCSI out}.

Given that the capacity of the  WFFD-RCSI is to within 5 bits from the capacity of the version with integer, peak-limited channel, we can now
manipulate the capacity expression as
\eas{
C
& = \max_{P_{\Xoo,U|S}}  I(\Yoo;U|A)-I(S;U) \\
%
%
& = \max_{P_{\Xoo,U|S}} I(\Yoo;X,S|A) -H(S)+H(S|\Yh,U,A) \\
& \leq \max_{P_{\Xoo,U|S}} I(\Yoo;X,S|A) -H(S)+H(S|\Yh,A),
\label{eq:p4 v2}
%
}
where \eqref{eq:p4 v2}  follows from the fact that $\Xoo=\Xoo(U,S)$, that is $\Xoo$  can be taken to be a deterministic function of $S$ and $U$.
The term $H(S|\Yh,U,A)$ can be bound analogously as in the proof of  \ref{th:NCSI out} in App. \ref{app:NCSI out}  and it can be easily verified that
 the condition for which $Q_S (\Yh)$ has cardinality zero corresponds to the condition in \eqref{eq:RCSI OUT conditions}.
%
%
%
%
%
%
%
%
%

\end{document}